\documentclass[amsmath, superscriptaddress, amssymb, aps, prd, longbibliography,nofootinbib,onecolumn, notitlepage,10pt]{revtex4-2}
\usepackage{amsmath,amssymb,mathtools}
\usepackage{natbib}
\usepackage[colorlinks=true,linkcolor=blue,citecolor=blue]{hyperref}
\usepackage{cleveref}
\usepackage[normalem]{ulem}
\usepackage{subcaption}

\usepackage{braket}
\usepackage{xcolor}
\usepackage{graphicx}
\usepackage{bm}
\labelformat{appendix}{Appendix #1}
\usepackage{comment}

\let\calccommentout\iffalse 
\let\calcshow\iftrue 

\crefname{section}{Section}{Sections}
\crefname{subsection}{Section}{Sections}
\crefname{subsubsection}{Section}{Sections}

\crefname{appendix}{Appendix}{Appendices}
\labelformat{equation}{Eq.~(#1)} 
\labelformat{figure}{Fig.~#1} 
\labelformat{subfigure}{Fig.~\thefigure#1} 
\labelformat{table}{Tab.~#1} 
\labelformat{appendix}{Appendix #1}

\begin{document}

\title{Asymptotic Decoherence of an Unruh--DeWitt Detector in de
Sitter Spacetime: Conformal versus Minimal Coupling}
\author{Kavitha A}
\email{kavitha.a2025@vitstudent.ac.in}
\author{Shagun Kaushal}
\email{shagun.kaushal@vit.ac.in}
\affiliation{Department of Physics, School of Advanced Sciences, Vellore Institute of Technology, Vellore 632014, India}

\begin{abstract}
We investigate the loss of coherence of a two-level Unruh--DeWitt
detector coupled to massless real scalar fields in
$(1+3)$-dimensional de Sitter spacetime. Treating the
detector--field interaction perturbatively to second order, we derive
analytic expressions for the real asymptotic coherence-decay
coefficients for conformally and minimally coupled scalar fields.
For the conformally coupled field, the coefficient remains finite in
the small-gap limit and grows linearly with the detector energy gap
in the large-gap regime. For the massless minimally coupled field,
the logarithmic sector of the Wightman function produces an additional
positive contribution. This provides an explicit analytic extension
of the previously known infrared enhancement of detector transition
responses to the asymptotic coefficient governing detector coherence.
The explicitly average-time-dependent sector does not contribute to
the real nonzero-frequency asymptotic coefficient under the
adiabatically regulated half-line prescription adopted here, when
the regulator is removed at fixed average time and fixed positive
detector gap. This asymptotic result does not describe the complete
finite-time dynamics, which depends on the switching function and
observation interval. Within the stated prescription, the coefficients satisfy
$\Gamma_{\rm MMC}>\Gamma_{\rm CC}$ for $\omega>0$, with
$\Gamma_{\rm MMC}/\Gamma_{\rm CC}
=1+(H/\omega)^2$.
The distinction is most pronounced in the small-gap regime, whereas
the two coefficients approach one another for large detector gaps.
\end{abstract}

\maketitle
\section{Introduction}
Quantum field theory in curved spacetime provides a framework for
studying quantum fields in the presence of gravitational backgrounds,
leading to phenomena such as the Unruh effect \cite{Unruh:1976db, UW, Crispino:2007eb} and Hawking
radiation \cite{Hawking, Hawking1}. These effects illustrate that the particle content
associated with a quantum field can depend on the observer and the
underlying spacetime geometry. A useful operational probe of such
field fluctuations is provided by the Unruh--DeWitt (UDW) detector,
originally introduced as a simple model of a point-like two-level
quantum system interacting locally with a quantum field along a
prescribed classical trajectory \cite{BirrellDavies1984, UDW, Fulling, Wald:1995yp, Fulling:1989nb,
Louko:2007mu}.

Traditionally, UDW detectors have been used to characterize
quantum-field fluctuations through transition probabilities and
response functions \cite{Louko:2007mu, Takagi:1986kn, Sriramkumar:1994pb,
Schlicht:2003iy, Hu:2004zu, Louko:2006zv, Hu:2015lda}. More recently, the open-quantum-system perspective
has provided a complementary framework in which the detector is
regarded as a quantum subsystem interacting with the field as its
environment \cite{JuarezAubry:2019, Tian:2015sda, Moustos:2016lol, Moustos:2016lol, Yu:2011eq, Breuer:2002pc, Lima:2023pyt, Anastopoulos:1999ht}. By tracing over the field degrees of freedom, one obtains
the reduced density matrix of the detector, whose off-diagonal
elements characterize quantum coherence between the detector energy
eigenstates \cite{Breuer:2002pc, Schlosshauer:2003}. Their perturbative evolution is determined by the field
correlation function sampled along the detector trajectory and
therefore provides an operational probe of how the quantum field and
the underlying spacetime affect the coherence of a localized quantum
system \cite{Zurek:1991vd, Koksma:2009wa, Nesterov:2020exl,
Foo:2020dzt, Kollas:2022wgj, Xu:2023tdt, Han:2025pql, Gundhi:2025bwj, Sahota:2026imu}. In the present work, we focus specifically on the real
asymptotic perturbative coefficient governing the secular change of
the coherence magnitude, rather than the complete switching-dependent
finite-time evolution. Beyond single-detector dynamics, UDW detectors
also provide an important operational framework for probing vacuum
correlations, particularly through entanglement harvesting
\cite{Barman:2022xht, Pozas-Kerstjens:2015gta, Tian:2014jha, Benatti:2004ee, Kaushal:2024zfi,
Nambu:2013rta, Sachs:2017exo, Liu:2023zro, Kukita:2017etu,
Bhattacharya:2022ahn, Martin-Martinez:2015qwa, Kaushal:2026ykl,
Reznik:2003mnx, Barman:2021kwg}.

In an expanding background such as de Sitter spacetime, quantum-field
fluctuations give rise to a thermal response for a comoving detector,
with the corresponding temperature set by the Gibbons--Hawking
temperature \cite{Yu:2011eq, GW, Zhou:2022jkg}. Detector response and
open-system dynamics in de Sitter spacetime have been studied in a
variety of settings, particularly for conformally coupled scalar
fields \cite{Garbrecht:2004du, Ali:2020gij, Allen:1987tz}. The
decoherence properties of an Unruh--DeWitt detector, however, can also
depend sensitively on the infrared structure of the field correlation
function. This becomes particularly relevant for a massless minimally
coupled (MMC) scalar field, whose two-point function differs
qualitatively from the conformally coupled case through the presence
of a logarithmic infrared contribution and an explicitly
de Sitter-breaking term
\cite{Allen:1985ux, Allen:1987tz, Miao:2010vs, Ford:1984hs}. Previous studies have established that conformally and minimally
coupled scalar fields produce different transition responses for
Unruh--DeWitt detectors in de Sitter spacetime, particularly in the
infrared regime \cite{Garbrecht:2004du,Ali:2020gij}. These transition
responses, however, do not by themselves determine the perturbative
evolution of the off-diagonal element of the detector density matrix.
It therefore remains important to determine how the different sectors
of the MMC correlation function enter the detector coherence kernel
and whether the resulting coherence-decay coefficient can be
distinguished from its conformally coupled counterpart.

In this work, we perform a direct analytic comparison of the asymptotic
loss of coherence of a two-level Unruh--DeWitt detector coupled to
massless conformally coupled and minimally coupled real scalar fields in
$(1+3)$-dimensional de Sitter spacetime. Starting from the
detector--field interaction and working perturbatively to second order
in the coupling, we extract the real asymptotic decoherence coefficient
governing the secular evolution of the off-diagonal element of the
detector density matrix.

Our contribution is an explicit analytic derivation of the real
asymptotic coherence-decay coefficients for massless conformally
coupled and minimally coupled real scalar fields. Although the
corresponding detector transition responses have been studied
previously, the coherence kernel involves a symmetric combination of
half-line Wightman transforms and therefore requires a separate
analysis. For the conformally coupled field, this coefficient is
determined by the stationary thermal correlation function. For the
MMC field, we decompose the Wightman function into its conformal
$1/y$ contribution, logarithmic relative-time contribution, and
explicitly average-time-dependent sector. The latter is
treated using an adiabatically regulated half-line transform. With the
regulator removed at fixed average time and fixed detector gap
$\omega>0$, this sector does not contribute to the real
nonzero-frequency asymptotic coefficient. This statement does not
describe the complete finite-time detector dynamics, for which the
switching function and average-time-dependence must be treated
simultaneously in the original double-time integral.

The logarithmic relative-time sector produces an additional positive
contribution to the real coherence-decay coefficient. Consequently,
within the weak-coupling, long-interaction, and secular prescription
employed here,
\begin{equation}
\Gamma_{\rm MMC}>\Gamma_{\rm CC},
\qquad
\omega>0.
\label{GammaInequalityIntro}
\end{equation}
The relative enhancement is
\begin{equation}
\frac{\Gamma_{\rm MMC}}{\Gamma_{\rm CC}}
=
1+\left(\frac{H}{\omega}\right)^2.
\label{GammaRatioIntro}
\end{equation}
The distinction between minimal and conformal coupling is most
pronounced when the detector gap is comparable to or smaller than the
de Sitter curvature scale. In this regime, the MMC coefficient
exhibits strong infrared enhancement, whereas the two coefficients
approach one another for $\omega/H\gg1$.

The remainder of this paper is organized as follows.
In Section~\ref{sec:setup}, we introduce the detector--field system
and derive the reduced detector dynamics to second order.
In Section~\ref{sec:conformally_coupled}, we obtain the real
asymptotic coefficient for the massless conformally coupled field.
Section~\ref{sec:mmc} is devoted to the massless minimally coupled
field, including the regulated treatment of its explicitly
average-time-dependent sector and the contribution generated by its
logarithmic structure. In Section~\ref{sec:conclusion}, we summarize
our results and discuss possible extensions. Technical details of the
Fourier-transform calculations are presented in
Appendices~\ref{app:conformally_coupled_residues} and~\ref{app:mmc}.
Throughout this work, we use natural units, $\hbar=c=1$.

\section{The Setup}
\label{sec:setup}

We consider a two-level Unruh--DeWitt (UDW) detector interacting
locally with a real scalar field in $(1+3)$-dimensional de Sitter
spacetime. The detector follows a prescribed classical trajectory
$x(\tau)$ parametrized by its proper time $\tau$. The total
Hamiltonian is
\begin{equation}
H
=
H_d+H_\phi+H_{\rm int}(\tau),
\label{H_total}
\end{equation}
where $H_d$, $H_\phi$, and $H_{\rm int}$ denote the free detector,
free field, and interaction Hamiltonians, respectively. The detector
Hamiltonian is
\begin{equation}
H_d
=
\frac{\omega}{2}\hat{\sigma}_z,
\end{equation}
where $\omega>0$ is the detector energy gap and $\hat{\sigma}_z$ is
the Pauli operator in the detector energy eigenbasis.

The detector--field interaction is described by
\begin{equation}
H_{\rm int}(\tau)
=
\lambda\,
\chi(\tau)\,
\hat{\sigma}_x\otimes\phi[x(\tau)],
\label{Hint}
\end{equation}
where $\lambda$ is the detector--field coupling strength and
$\chi(\tau)$ is a real switching function. We set the initial
interaction time to $\tau_i=0$ and denote the final interaction time
by $\tau_f$. The switching function is assumed to possess a long
constant plateau with smooth adiabatic turn-on and turn-off regions.
The asymptotic coefficients studied below are defined by taking
\begin{equation}
\tau_f\longrightarrow\infty,
\qquad
\omega>0\ \text{fixed},
\label{long_interaction_limit}
\end{equation}
with the observation point lying within the bulk of the plateau.
Following Ref.~\cite{Ali:2020gij}, the infinite relative-time
transforms are understood in the adiabatically regulated sense, with
the regulator removed after integration. Finite switching regions
may generate transient boundary contributions, but these are not part
of the asymptotic secular coefficient considered here.

The detector monopole operator is
\begin{equation}
\hat{\sigma}_x
=
|g\rangle\langle e|
+
|e\rangle\langle g|,
\end{equation}
and $\phi[x(\tau)]$ denotes the scalar field evaluated along the
detector trajectory.

We assume that the field is initially in the vacuum state $|0\rangle$
and that the detector is prepared in the pure state
\begin{equation}
|\psi\rangle
=
\sqrt{1-p}\,|g\rangle
+
\sqrt{p}\,|e\rangle,
\qquad
p\in[0,1].
\label{detectorstate}
\end{equation}
In the basis $\{|e\rangle,|g\rangle\}$, the corresponding detector
density matrix is
\begin{equation}
\rho_d(0)
=
\begin{pmatrix}
p & \sqrt{p(1-p)}
\\
\sqrt{p(1-p)} & 1-p
\end{pmatrix}.
\label{initialDM}
\end{equation}
Assuming that the detector and field are initially uncorrelated, the
total initial state is
\begin{equation}
\rho(0)
=
\rho_d(0)\otimes|0\rangle\langle0|.
\label{rho_total}
\end{equation}

To determine the reduced detector dynamics, we work in the interaction
picture. The detector monopole operator becomes
\begin{equation}
\hat{\mu}(\tau)
=
\hat{\sigma}_{+}e^{i\omega\tau}
+
\hat{\sigma}_{-}e^{-i\omega\tau},
\label{mu}
\end{equation}
where
\begin{equation}
\hat{\sigma}_{+}
=
|e\rangle\langle g|,
\qquad
\hat{\sigma}_{-}
=
|g\rangle\langle e|.
\end{equation}
The interaction-picture Hamiltonian is therefore
\begin{equation}
H_{\rm int}^{I}(\tau)
=
\lambda\,
\chi(\tau)\,
\hat{\mu}(\tau)\otimes\Phi^I(\tau),
\label{Hint_t}
\end{equation}
where $\Phi^I(\tau)$ denotes the interaction-picture scalar field
evaluated along the detector trajectory.

At the final interaction time $\tau_f$, the total state is
\begin{equation}
\rho^I(\tau_f)
=
U_{\rm int}^{I}(\tau_f)\,
\rho(0)\,
U_{\rm int}^{I\dagger}(\tau_f),
\label{rho_evolution}
\end{equation}
where
\begin{equation}
U_{\rm int}^{I}(\tau_f)
=
\mathcal T
\exp\left[
-i\int_0^{\tau_f} d\tau_1\,
H_{\rm int}^{I}(\tau_1)
\right].
\label{Uint}
\end{equation}
Expanding perturbatively to second order in $\lambda$ gives
\begin{align}
\rho^I(\tau_f)
={}&
\rho(0)
-i\int_0^{\tau_f} d\tau_1\,
\left[
H_{\rm int}^{I}(\tau_1),
\rho(0)
\right]
\nonumber\\
&-
\int_0^{\tau_f} d\tau_1
\int_0^{\tau_f} d\tau_2\,
\Theta(\tau_1-\tau_2)
\left[
H_{\rm int}^{I}(\tau_1),
\left[
H_{\rm int}^{I}(\tau_2),
\rho(0)
\right]
\right]
+
\mathcal O(\lambda^3).
\label{rho_total_int}
\end{align}

The reduced detector density matrix is obtained by tracing over the
field degrees of freedom:
\begin{equation}
\rho_d^I(\tau_f)
=
\operatorname{Tr}_{\phi}
\left[
\rho^I(\tau_f)
\right].
\label{rhod}
\end{equation}
Because the interaction Hamiltonian is linear in the field and
$\langle0|\phi(x)|0\rangle=0$, the first-order contribution vanishes.
The leading nontrivial correction therefore appears at order
$\lambda^2$ and is determined by the positive-frequency Wightman
function evaluated along the detector trajectory:
\begin{equation}
iG^+(\tau_1,\tau_2)
\equiv
\langle0|
\phi[x(\tau_1)]
\phi[x(\tau_2)]
|0\rangle.
\label{Gdplus}
\end{equation}
The factor $i$ in the notation $iG^+$ follows the propagator
convention used in Refs.~\cite{Ali:2020gij,Xu:2023tdt}; the
expectation value on the right-hand side is the Wightman two-point
function. For the free Gaussian field state considered here, odd
field correlation functions vanish, so the next nonzero contribution
to the reduced detector state occurs at order $\lambda^4$.

Tracing the second-order contribution in~\ref{rho_total_int} over the field gives
\begin{align}
\rho_d^{I(2)}(\tau_f)
={}&
-\lambda^2
\int_0^{\tau_f} d\tau_1
\int_0^{\tau_f} d\tau_2\,
\Theta(\tau_1-\tau_2)\,
\chi(\tau_1)\chi(\tau_2)
\nonumber\\
&\times
\Big\{
iG^+(\tau_1,\tau_2)
\big[
\hat{\mu}(\tau_1),
\hat{\mu}(\tau_2)\rho_d(0)
\big]
\nonumber\\
&\hspace{1.2cm}
+
iG^+(\tau_2,\tau_1)
\big[
\rho_d(0)\hat{\mu}(\tau_2),
\hat{\mu}(\tau_1)
\big]
\Big\}.
\label{reduced_second_order}
\end{align}

We introduce the relative and average internal proper times
\begin{equation}
\Delta\tau
=
\tau_1-\tau_2,
\qquad
T
=
\frac{\tau_1+\tau_2}{2},
\label{relative_average_time}
\end{equation}
and write
\begin{equation}
iG^+(\tau_1,\tau_2)
=
iG^+(\Delta\tau,T).
\label{WightmanRelativeAverage}
\end{equation}
Here, $T$ is the internal average-time coordinate and must be
distinguished from the final interaction time $\tau_f$. This distinction is
particularly important for the massless minimally coupled field,
whose Wightman function depends explicitly on $T$.

At finite $\tau_f$, the integration domains of $T$ and $\Delta\tau$ are
correlated. In particular,
\begin{equation}
0\leq T\leq \tau_f,
\qquad
-2\min(T,\tau_f-T)
\leq
\Delta\tau
\leq
2\min(T,\tau_f-T).
\label{finite_time_domain}
\end{equation}
The switching functions in~\ref{reduced_second_order} therefore
contribute explicitly near the boundaries of this domain. The
complete finite-time evolution cannot consequently be obtained by
independently extending the relative-time integration limits to
infinity.

In the present work, we instead extract the secular bulk contribution
in the weak-coupling, long-interaction regime defined in~\ref{long_interaction_limit}. Substitution of~\ref{mu}
into~\ref{reduced_second_order} separates the terms involving
the relative-time phases $e^{\pm i\omega\Delta\tau}$ from the
counter-rotating terms proportional to
\begin{equation}
e^{\pm i\omega(\tau_1+\tau_2)}
=
e^{\pm2i\omega T}.
\end{equation}
The counter-rotating terms are removed by coarse-graining over
timescales large compared with $\omega^{-1}$. Within the bulk of the
long constant plateau, the relative-time memory integrals may then be
extended to the corresponding half-lines. These steps constitute the
weak-coupling, long-interaction, and secular prescription employed here
\cite{Kaplanek:2019dqu, Spohn:1980zz}. They are not intended to describe the
complete switching-dependent finite-time evolution.

Following Ref.~\cite{Xu:2023tdt}, we focus on the off-diagonal element of the reduced detector density matrix. After evaluating the $(1,2)$ matrix element, the terms proportional to $\rho_{21}^{I}(0)e^{\pm 2i\omega T}$ are identified as counter-rotating contributions and are neglected under the secular coarse-graining. The remaining terms are proportional to $\rho_{12}^{I}(0)$ and can be expressed in terms of the asymptotic coherence kernel
\begin{align}
\mathcal D(\omega,T)
=\int_0^\infty
d(\Delta\tau)\,
iG^+(\Delta\tau,T)e^{i\omega\Delta\tau}
+
\int_{-\infty}^{0}
d(\Delta\tau)\,
iG^+(\Delta\tau,T)e^{-i\omega\Delta\tau}.
\label{decoherence_functional}
\end{align}
The infinite integrals in~\ref{decoherence_functional} are understood in the regulated sense specified below~\ref{long_interaction_limit}.

Within this coarse-grained bulk description, $T$ serves as the central-time evolution parameter. The secular part of the second-order evolution of the detector coherence is therefore
\begin{equation}
\left.
\frac{\partial\rho_{12}^{I}(T)}{\partial T}
\right|_{\rm sec}
=
-\lambda^2
\rho_{12}^{I}(0)
\mathcal D(\omega,T)
+
\mathcal O(\lambda^4),
\label{coherence_eom}
\end{equation}
where
\begin{equation}
\rho_{12}^{I}(0)=\sqrt{p(1-p)}
\end{equation}
is the initial detector coherence. \ref{coherence_eom} describes the secular bulk contribution obtained after implementing the weak-coupling, long-interaction, and secular prescription. In particular, it should not be interpreted as resulting from an identification of the internal average-time coordinate $T$ with the finite upper interaction time $\tau_f$ appearing in \ref{reduced_second_order}.

In general, $\mathcal D(\omega,T)$ is complex. To second order in the
coupling, its real part governs the change in the magnitude of the
detector coherence, whereas its imaginary part governs the phase. We
therefore define the real asymptotic perturbative decoherence
coefficient by
\begin{equation}
\Gamma(\omega,T)
=
\lambda^2
\operatorname{Re}\mathcal D(\omega,T).
\label{decoherence_rate}
\end{equation}
Accordingly,
\begin{equation}
\left.
\frac{\partial|\rho_{12}^{I}(T)|}{\partial T}
\right|_{\rm sec}
=
-\Gamma(\omega,T)
|\rho_{12}^{I}(0)|
+
\mathcal O(\lambda^4).
\label{coherence_magnitude_evolution}
\end{equation}
This is a second-order perturbative relation involving the initial
coherence and should not be interpreted as an exact exponential decay
law. Obtaining reliable late-time predictions generally requires an
appropriate resummation or a controlled time-local master-equation
treatment Ref.~\cite{Kaplanek:2019dqu}.

For correlation functions that are stationary along the detector
trajectory,
\begin{equation}
iG^+(\tau_1,\tau_2)
=
iG^+(\Delta\tau),
\end{equation}
the kernel becomes independent of the average-time:
\begin{equation}
\mathcal D(\omega,T)
=
\mathcal D(\omega).
\end{equation}
The corresponding asymptotic coefficient is
\begin{equation}
\Gamma(\omega)
=
\lambda^2
\operatorname{Re}\mathcal D(\omega).
\label{stationary_decoherence_rate}
\end{equation}
The conformally coupled field considered below belongs to this
stationary class. For the massless minimally coupled field, we retain
the explicit average-time-dependence before applying the regulated
nonzero-frequency prescription. The resulting quantity should be
understood as an asymptotic spectral coefficient rather than as the
complete finite-time coherence evolution.

In the following sections, we evaluate the real asymptotic
perturbative decoherence coefficients for massless conformally and
minimally coupled scalar fields in de Sitter spacetime.

\section{The Conformally Coupled Scalar Field}
\label{sec:conformally_coupled}

We first consider a massless conformally coupled real scalar field in
$(1+3)$-dimensional de Sitter spacetime. Along a comoving detector
trajectory, the positive-frequency Wightman function is stationary
and depends only on the proper-time separation $\Delta\tau$. It is
given by Ref.~\cite{Ali:2020gij}
\begin{equation}
iG^+_{\rm CC}(\Delta\tau)
=
-\frac{H^2}{16\pi^2}
\frac{1}{
\sinh^2\!\left[
\frac{H}{2}(\Delta\tau-i\epsilon)
\right]
},
\label{GCC}
\end{equation}
where $H$ is the de Sitter Hubble parameter and
$\epsilon\rightarrow0^+$ implements the standard positive-frequency
prescription. Since ~\ref{GCC} is independent of the average
proper time $T$, the coherence kernel defined in
~\ref{decoherence_functional} is stationary and may be denoted by
$\mathcal D_{\rm CC}(\omega)$.

Substituting~\ref{GCC} into~\ref{decoherence_functional}, we obtain
\begin{align}
\mathcal D_{\rm CC}(\omega)
=
-\frac{H^2}{16\pi^2}
\int_0^\infty d(\Delta\tau)\,
\frac{e^{i\omega\Delta\tau}}
{\sinh^2\!\left[
\frac{H}{2}(\Delta\tau-i\epsilon)
\right]}
-
\frac{H^2}{16\pi^2}
\int_{-\infty}^{0}d(\Delta\tau)\,
\frac{e^{-i\omega\Delta\tau}}
{\sinh^2\!\left[
\frac{H}{2}(\Delta\tau-i\epsilon)
\right]}.
\label{DCC_integral}
\end{align}
Introducing $\tau=-\Delta\tau$ in the second integral and using
$\sinh(-x)=-\sinh x$,~\ref{DCC_integral} becomes
\begin{align}
\mathcal D_{\rm CC}(\omega)
=
-\frac{H^2}{16\pi^2}
\int_0^\infty d\tau\,
e^{i\omega\tau}
\left[
\frac{1}{
\sinh^2\!\left[
\frac{H}{2}(\tau-i\epsilon)
\right]}
+
\frac{1}{
\sinh^2\!\left[
\frac{H}{2}(\tau+i\epsilon)
\right]}
\right].
\label{DCC_symmetric}
\end{align}

The real part of this half-line kernel can be related directly to the
full-line response spectrum
\begin{equation}
\mathcal F_{\rm CC}(\omega)
=
\int_{-\infty}^{\infty}
d(\Delta\tau)\,
iG^+_{\rm CC}(\Delta\tau)
e^{i\omega\Delta\tau}.
\label{FCC}
\end{equation}
Using the hermiticity property
$iG^+_{\rm CC}(-\Delta\tau)
=[iG^+_{\rm CC}(\Delta\tau)]^*$, one finds
\begin{equation}
\operatorname{Re}\mathcal D_{\rm CC}(\omega)
=
\frac{1}{2}
\left[
\mathcal F_{\rm CC}(\omega)
+
\mathcal F_{\rm CC}(-\omega)
\right].
\label{ReDCC_response}
\end{equation}
Thus, the real part of the coherence kernel is determined by the
symmetric combination of the positive- and negative-frequency
response spectra.

For a fixed positive detector gap, $\omega>0$, the full-line Fourier
transforms can be evaluated analytically and give
\begin{equation}
\mathcal F_{\rm CC}(\omega)
=
\frac{\omega}{2\pi}
\frac{1}{1-e^{-2\pi\omega/H}},
\label{RCCpositive}
\end{equation}
and
\begin{equation}
\mathcal F_{\rm CC}(-\omega)
=
\frac{\omega}{2\pi}
\frac{1}{e^{2\pi\omega/H}-1}.
\label{RCCnegative}
\end{equation}
The contour evaluation of these Fourier transforms is presented in
Appendix~\ref{app:conformally_coupled_residues}.

Substitution of~\ref{RCCpositive} and
\ref{RCCnegative} into~\ref{ReDCC_response} yields
\begin{align}
\operatorname{Re}\mathcal D_{\rm CC}(\omega)
&=
\frac{\omega}{4\pi}
\left[
\frac{1}{1-e^{-2\pi\omega/H}}
+
\frac{1}{e^{2\pi\omega/H}-1}
\right]
\nonumber\\
&=
\frac{\omega}{4\pi}
\coth\left(
\frac{\pi\omega}{H}
\right).
\label{ReDCC_result}
\end{align}
Consequently, the real asymptotic perturbative decoherence coefficient
defined in~\ref{decoherence_rate} is
\begin{equation}
\Gamma_{\rm CC}(\omega)
=
\lambda^2
\frac{\omega}{4\pi}
\coth\left(
\frac{\pi\omega}{H}
\right).
\label{GammaCCdimensional}
\end{equation}
Equivalently, in units of the Hubble parameter,
\begin{equation}
\frac{\Gamma_{\rm CC}}{H}
=
\lambda^2
\frac{\omega}{4\pi H}
\coth\left(
\frac{\pi\omega}{H}
\right).
\label{GammaCC}
\end{equation}
~\ref{GammaCC} shows that the dimensionless coefficient is
controlled by the ratio $\omega/H$ and scales quadratically with the
detector--field coupling $\lambda$.

Its limiting behavior is particularly simple. In the small-gap
regime, $\omega/H\ll1$, the expansion
\begin{equation}
\coth x
=
\frac{1}{x}
+
\mathcal O(x),
\qquad
x\rightarrow0,
\end{equation}
gives
\begin{equation}
\frac{\Gamma_{\rm CC}}{H}
\longrightarrow
\frac{\lambda^2}{4\pi^2},
\qquad
\frac{\omega}{H}\rightarrow0^+.
\label{GammaCCsmallgap}
\end{equation}
Thus, although the detector gap approaches zero from above, the
dimensionless conformally coupled coefficient remains finite.

In the opposite regime, $\omega/H\gg1$,
\begin{equation}
\coth\left(
\frac{\pi\omega}{H}
\right)
\longrightarrow1,
\end{equation}
and therefore
\begin{equation}
\frac{\Gamma_{\rm CC}}{H}
\simeq
\lambda^2
\frac{\omega}{4\pi H}.
\label{GammaCClargegap}
\end{equation}
The asymptotic coefficient is therefore linear in $\omega/H$ at large
detector gaps.

\begin{figure}
    \centering
    \includegraphics[width=0.5\linewidth]{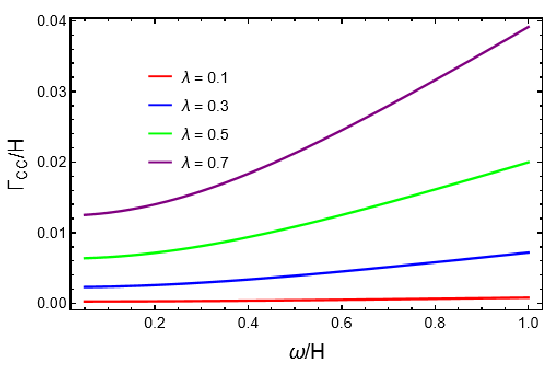}
   \caption{\it Dimensionless real asymptotic perturbative decoherence
coefficient $\Gamma_{\rm CC}/H$ for a two-level UDW detector coupled
to a massless conformally coupled scalar field in
$(1+3)$-dimensional de Sitter spacetime.}
    \label{fig:CC}
\end{figure}

\ref{fig:CC} illustrates these features for representative
values of the detector--field coupling. The coefficient approaches
the finite value $\lambda^2/(4\pi^2)$ in the small-gap limit and
becomes approximately linear in $\omega/H$ in the large-gap regime.

%%%%%%%%%%%%%%%%%%%%%%%%%%%%%%%%%%%%%%%%%%%%%%%%%%%%%%%%%%%%%%%%%%%%%%%%%%

\section{Massless Minimally Coupled Scalar Field}
\label{sec:mmc}

We now consider a massless minimally coupled (MMC) real scalar field
in $(1+3)$-dimensional de Sitter spacetime. Unlike the conformally
coupled case, the MMC Wightman function contains a logarithmic
contribution associated with the infrared structure of the field and
an explicitly average-time-dependent term that breaks the full de
Sitter invariance. This average-time-dependence must be retained until
the real asymptotic coefficient is extracted.

For the massless minimally coupled scalar field, we use the de Sitter-breaking two-point function employed in Ref.~\cite{Ali:2020gij}. It defines an $E(3)$-invariant state in the spatially flat Poincar\'e patch. Here, $E(3)$ is the symmetry group generated by spatial translations and rotations. Thus, although the state does not preserve the complete de Sitter symmetry, it remains homogeneous and isotropic on the spatial hypersurfaces.

Adopting the infrared convention of Ref.~\cite{Ali:2020gij}, the positive-frequency Wightman function evaluated along the worldline of a comoving detector takes the form
\begin{equation}
iG_{\rm MMC}^{+}(\Delta\tau,T)
=
\frac{H^{2}}{4\pi^{2}}
\left[
\frac{1}{y(\Delta\tau)}
-\frac{1}{2}\ln y(\Delta\tau)
+\frac{1}{2}\ln\bigl(a(\tau_1)a(\tau_2)\bigr)
+\ln 2-\frac{1}{4}
\right],
\label{MMCWF}
\end{equation}
where
\begin{equation}
y(\Delta\tau)
=
-4\sinh^{2}\left[
\frac{H}{2}(\Delta\tau-i\epsilon)
\right],
\qquad
\epsilon\rightarrow0^{+}.
\label{y}
\end{equation}

The infrared behavior of an exactly massless minimally coupled scalar field requires particular care. Its zero mode prevents the construction of an ordinary normalizable Fock vacuum that is invariant under the full de Sitter group, as discussed in Refs.~\cite{Allen:1985ux,Allen:1987tz}. Consequently, both the choice of state and the associated infrared prescription form part of the physical specification of the problem.

A different choice of the additive infrared constant changes only the part of the two-point function that is independent of the relative time $\Delta\tau$. Such a modification can influence contributions supported at zero frequency, as well as switching-dependent transients at finite interaction times. Under the regulated asymptotic prescription used in the present analysis, however, it does not modify the real nonzero-frequency coefficient when the detector gap is held fixed at $\omega>0$.

Unlike the conformally coupled correlation function,~\ref{MMCWF} is not solely a function of the relative proper time
$\Delta\tau$. The general coherence kernel defined in~\ref{decoherence_functional} must therefore initially be
retained in the form
\begin{align}
\mathcal D_{\rm MMC}(\omega,T)
=
\int_{0}^{\infty}
d(\Delta\tau)\,
iG_{\rm MMC}^{+}(\Delta\tau,T)
e^{i\omega\Delta\tau}
+
\int_{-\infty}^{0}
d(\Delta\tau)\,
iG_{\rm MMC}^{+}(\Delta\tau,T)
e^{-i\omega\Delta\tau}.
\label{DMMCgeneral}
\end{align}
Transforming the second integral according to
$\Delta\tau\rightarrow-\Delta\tau$ gives
\begin{equation}
\mathcal D_{\rm MMC}(\omega,T)
=
\int_{0}^{\infty}
d(\Delta\tau)\,
e^{i\omega\Delta\tau}
\left[
iG_{\rm MMC}^{+}(\Delta\tau,T)
+
iG_{\rm MMC}^{+}(-\Delta\tau,T)
\right].
\label{MMCsym}
\end{equation}
The symmetric Wightman combination is
\begin{align}
&iG_{\rm MMC}^{+}(\Delta\tau,T)
+iG_{\rm MMC}^{+}(-\Delta\tau,T)
\nonumber\\
&\qquad=
\frac{H^{2}}{4\pi^{2}}
\Bigg[
\frac{1}{y(\Delta\tau)}
+\frac{1}{y(-\Delta\tau)}
-\frac{1}{2}
\left(
\ln y(\Delta\tau)
+
\ln y(-\Delta\tau)
\right)
+2C(T)
\Bigg],
\label{MMCsymmetric}
\end{align}
where
\begin{equation}
C(T)
=
\frac{1}{2}
\ln\!\bigl[a(\tau_1)a(\tau_2)\bigr]
+\ln2-\frac{1}{4}.
\label{MMCconstant}
\end{equation}

For the comoving detector, proper time coincides with cosmological
time and
\begin{equation}
a(\tau)=e^{H\tau}.
\end{equation}
Since
\begin{equation}
T=\frac{\tau_1+\tau_2}{2},
\end{equation}
we have
\begin{equation}
\frac{1}{2}
\ln\!\left[a(\tau_1)a(\tau_2)\right]
=
\frac{H}{2}(\tau_1+\tau_2)
=
HT,
\end{equation}
and hence
\begin{equation}
C(T)
=
HT+\ln2-\frac14.
\label{CT}
\end{equation}
The MMC coherence kernel consequently decomposes as
\begin{equation}
\mathcal D_{\rm MMC}(\omega,T)
=
\mathcal D_{1/y}(\omega)
+
\mathcal D_{\log}(\omega)
+
\mathcal D_C(\omega,T).
\label{DMMCdecomposition}
\end{equation}
We now evaluate these three contributions separately.

\subsection{The $1/y$ contribution}

The first term in ~\ref{MMCWF} is
\begin{equation}
\frac{H^2}{4\pi^2}\frac{1}{y(\Delta\tau)}
=
-\frac{H^2}{16\pi^2}
\frac{1}{
\sinh^2\!\left[
\frac{H}{2}(\Delta\tau-i\epsilon)
\right]}.
\label{MMC_conformal_sector}
\end{equation}
which is precisely the conformally coupled Wightman function given in~\ref{GCC}. Its contribution to the real part of the coherence
kernel is therefore
\begin{equation}
\operatorname{Re}\mathcal D_{1/y}(\omega)
=
\frac{\omega}{4\pi}
\coth\left(
\frac{\pi\omega}{H}
\right).
\label{ReDMMC1y}
\end{equation}

\subsection{Average-time-dependent contribution}
\label{sec:regulated_CT}

The explicitly average-time-dependent part of~\ref{MMCsymmetric} gives
\begin{equation}
\mathcal D_C(\omega,T)
=
\frac{H^2C(T)}{2\pi^2}
\int_0^\infty
d(\Delta\tau)\,
e^{i\omega\Delta\tau}.
\label{DC}
\end{equation}
In accordance with the adiabatic prescription specified in
Section~\ref{sec:setup}, we define the half-line transform by
\begin{equation}
\mathcal D_C^{(\eta)}(\omega,T)
=
\frac{H^2C(T)}{2\pi^2}
\int_0^\infty
d(\Delta\tau)\,
e^{(i\omega-\eta)\Delta\tau},
\qquad
\eta>0.
\label{DCregulated}
\end{equation}
The integral is elementary:
\begin{equation}
\mathcal D_C^{(\eta)}(\omega,T)
=
\frac{H^2C(T)}{2\pi^2}
\frac{1}{\eta-i\omega}.
\label{DCregulatedresult}
\end{equation}
Its real part is therefore
\begin{equation}
\operatorname{Re}\mathcal D_C^{(\eta)}(\omega,T)
=
\frac{H^2C(T)}{2\pi^2}
\frac{\eta}{\eta^2+\omega^2}
\longrightarrow0,
\qquad
\eta\rightarrow0^+,
\quad
\omega>0.
\label{ReDCzero}
\end{equation}

Equivalently, the unregulated half-line transform may be written as
the distributional identity
\begin{equation}
\int_0^\infty
d(\Delta\tau)\,
e^{i\omega\Delta\tau}
=
\pi\delta(\omega)
+
i\,\mathcal P\!\left(\frac{1}{\omega}\right),
\label{halflineidentity}
\end{equation}
where $\mathcal P$ denotes the Cauchy principal value. Thus the real
part of its regulated transform is supported only at zero frequency. This is
the half-line counterpart of the full-line detector-response
prescription used in Ref.~\cite{Ali:2020gij}, where the corresponding
de Sitter-breaking term is proportional to $\delta(\omega)$ and is
absent for a detector with a discrete nonzero energy gap.

It follows that
\begin{equation}
\operatorname{Re}\mathcal D_C(\omega,T)=0,
\qquad
\omega>0.
\label{ReDC}
\end{equation}
The complete MMC kernel still contains the imaginary contribution
$H^2C(T)/(2\pi^2\omega)$, which affects the phase of the detector
coherence. However, $C(T)$ does not contribute to the real
nonzero-frequency asymptotic coefficient defined in this work.

\subsubsection{Finite-duration interpretation}
\label{sec:finite_time_check}

The vanishing of the $C(T)$ contribution established above is an
asymptotic statement at fixed $\omega>0$. It does not imply that this
sector is absent from the complete finite-duration detector dynamics.
For example, if the relative-time integral is truncated at an upper
limit $s_{\max}$, then
\begin{equation}
\mathcal D_C^{(s_{\max})}(\omega,T)
=
\frac{H^2 C(T)}{2\pi^2}
\frac{e^{i\omega s_{\max}}-1}{i\omega},
\end{equation}
and hence
\begin{equation}
\operatorname{Re}\mathcal D_C^{(s_{\max})}(\omega,T)
=
\frac{H^2 C(T)}{2\pi^2}
\frac{\sin(\omega s_{\max})}{\omega}.
\end{equation}
This finite-duration contribution is oscillatory and depends on the
integration boundary. It is therefore not the secular coefficient
extracted in the adiabatic long-interaction limit. In the full-line
formulation, the corresponding relative-time-independent term is
supported only at zero frequency through a contribution proportional
to $\delta(\omega)$, as in Ref.~\cite{Ali:2020gij}. It consequently
does not contribute to the real asymptotic coefficient at fixed
$\omega>0$.

A complete finite-time calculation would require a specified
switching function and the simultaneous integration over the
correlated average- and relative-time domains. Such an analysis lies
beyond the scope of the present work.

\subsection{Logarithmic contribution}

The remaining contribution originates from the logarithmic term:
\begin{equation}
\mathcal D_{\log}(\omega)
=
-\frac{H^{2}}{8\pi^{2}}
\int_{0}^{\infty}
d(\Delta\tau)\,
e^{i\omega\Delta\tau}
\left[
\ln y(\Delta\tau)
+
\ln y(-\Delta\tau)
\right].
\label{Dlog1}
\end{equation}
Introducing the dimensionless variables
\begin{equation}
u=\frac{H\Delta\tau}{2},
\qquad
\Omega=\frac{2\omega}{H},
\label{dimensionless}
\end{equation}
the symmetric logarithmic combination becomes
\begin{equation}
\ln y(\Delta\tau)
+
\ln y(-\Delta\tau)
=
4u
+
4\ln\left(1-e^{-2u}\right),
\label{logcombination}
\end{equation}
where the branch-dependent imaginary parts cancel between the two
logarithms.~\ref{Dlog1} therefore reduces to
\begin{equation}
\mathcal D_{\log}(\omega)
=
-\frac{H}{\pi^{2}}
\int_{0}^{\infty}
du\,
e^{i\Omega u}
\left[
u+\ln\left(1-e^{-2u}\right)
\right].
\label{Dlog2}
\end{equation}
Both terms in \ref{Dlog2} must be retained. Their regulated
evaluation is presented in Appendix~\ref{app:mmc}. For a fixed
positive detector gap, the result is
\begin{equation}
\mathcal D_{\log}(\omega)
=
\frac{H^{3}}{4\pi^{2}\omega^{2}}
+
\frac{iH^{2}}{2\pi^{2}\omega}
\left[
\psi\!\left(
1-\frac{i\omega}{H}
\right)
+\gamma
\right],
\qquad
\omega>0,
\label{Dlogomega}
\end{equation}
where $\psi(z)$ is the digamma function and $\gamma$ is the
Euler--Mascheroni constant.

Using the identity
\begin{equation}
\operatorname{Im}\psi(1-ix)
=
-\frac{1}{2}
\left[
\pi\coth(\pi x)
-\frac{1}{x}
\right],
\qquad
x>0,
\label{digammaidentity}
\end{equation}
we find
\begin{align}
\operatorname{Re}\mathcal D_{\log}(\omega)
&=
\frac{H^3}{4\pi^2\omega^2}
-
\frac{H^2}{2\pi^2\omega}
\operatorname{Im}\psi\!\left(
1-\frac{i\omega}{H}
\right)
\nonumber\\
&=
\frac{H^3}{4\pi^2\omega^2}
+
\frac{H^2}{4\pi^2\omega}
\left[
\pi\coth\left(
\frac{\pi\omega}{H}
\right)
-\frac{H}{\omega}
\right].
\label{ReDlogintermediate}
\end{align}
The terms proportional to $H^3/\omega^2$ cancel exactly, leaving
\begin{equation}
\operatorname{Re}\mathcal D_{\log}(\omega)
=
\frac{H^{2}}{4\pi\omega}
\coth\left(
\frac{\pi\omega}{H}
\right).
\label{ReDlogsimplified}
\end{equation}
This cancellation is essential: the final infrared enhancement is a
property of the complete logarithmic contribution and cannot be
assigned to either term in~\ref{Dlog2} separately.

\subsection{MMC decoherence coefficient}

For a fixed positive detector gap, the explicitly average-time-dependent sector does not contribute to the real part of the
asymptotic kernel. Combining~\ref{ReDMMC1y},
\ref{ReDC}, and \ref{ReDlogsimplified}, we obtain
\begin{equation}
\operatorname{Re}\mathcal D_{\rm MMC}(\omega)
=
\left(
\frac{\omega}{4\pi}
+
\frac{H^{2}}{4\pi\omega}
\right)
\coth\left(
\frac{\pi\omega}{H}
\right),
\qquad
\omega>0.
\label{ReDMMCfinal}
\end{equation}
Using~\ref{decoherence_rate}, the corresponding real asymptotic
perturbative decoherence coefficient is
\begin{equation}
\Gamma_{\rm MMC}(\omega)
=
\lambda^{2}
\left(
\frac{\omega}{4\pi}
+
\frac{H^{2}}{4\pi\omega}
\right)
\coth\left(
\frac{\pi\omega}{H}
\right).
\label{GammaMMCdimensional}
\end{equation}
Equivalently,
\begin{equation}
\frac{\Gamma_{\rm MMC}}{H}
=
\lambda^{2}
\left[
\frac{\omega}{4\pi H}
+
\frac{H}{4\pi\omega}
\right]
\coth\left(
\frac{\pi\omega}{H}
\right).
\label{GammaMMCfinal}
\end{equation}

Comparison with the conformally coupled result in~\ref{GammaCC} gives
\begin{equation}
\frac{\Gamma_{\rm MMC}-\Gamma_{\rm CC}}{H}
=
\lambda^{2}
\frac{H}{4\pi\omega}
\coth\left(
\frac{\pi\omega}{H}
\right)
>0,
\qquad
\omega>0.
\label{GammaDifference}
\end{equation}
Thus, within the weak-coupling, long-interaction, and secular
prescription employed here, the MMC coefficient exceeds the
conformally coupled coefficient for every finite positive detector
gap. The enhancement is generated entirely by the logarithmic sector
of the MMC Wightman function.

The relative enhancement is particularly transparent in the ratio
\begin{equation}
\frac{\Gamma_{\rm MMC}}{\Gamma_{\rm CC}}
=
1+
\left(
\frac{H}{\omega}
\right)^2.
\label{GammaRatio}
\end{equation}
The coupling $\lambda$ cancels from this ratio. Minimal coupling is
therefore most consequential when the detector gap is comparable to
or smaller than the de Sitter curvature scale, whereas the two
coefficients approach one another when $\omega/H\gg1$.

In the small-gap regime, $\omega/H\ll1$,
\begin{equation}
\coth\left(
\frac{\pi\omega}{H}
\right)
=
\frac{H}{\pi\omega}
+
\frac{\pi\omega}{3H}
+
\mathcal O\!\left(
\frac{\omega^3}{H^3}
\right),
\end{equation}
and hence
\begin{equation}
\frac{\Gamma_{\rm MMC}-\Gamma_{\rm CC}}{H}
=
\frac{\lambda^2}{4\pi^2}
\left(
\frac{H}{\omega}
\right)^2
+
\frac{\lambda^2}{12}
+
\mathcal O\!\left(
\frac{\omega^2}{H^2}
\right).
\label{GammaDifferenceSmall}
\end{equation}
The MMC coefficient correspondingly has the small-gap expansion
\begin{equation}
\frac{\Gamma_{\rm MMC}}{H}
=
\frac{\lambda^2}{4\pi^2}
\left(
\frac{H}{\omega}
\right)^2
+
\lambda^2
\left(
\frac{1}{4\pi^2}
+
\frac{1}{12}
\right)
+
\mathcal O\!\left(
\frac{\omega^2}{H^2}
\right),
\qquad
\frac{\omega}{H}\ll1.
\label{GammaMMCsmall}
\end{equation}
Whereas the conformally coupled coefficient approaches the finite
value $\lambda^2/(4\pi^2)$, the MMC result exhibits a pronounced
infrared enhancement as $\omega/H\rightarrow0^+$. The divergence of
the second-order expression should not be interpreted as an
arbitrarily large physical decoherence rate. Instead, it indicates
that the infrared contribution increasingly dominates the
perturbative coefficient and that fixed-order perturbation theory
eventually loses quantitative reliability sufficiently close to the
degenerate limit. At an observation duration $\tau_{\rm obs}$,
perturbative consistency requires schematically
\begin{equation}
\Gamma_{\rm MMC}\,\tau_{\rm obs}\ll1.
\label{perturbativevalidity}
\end{equation}
The exactly degenerate case, $\omega=0$, is therefore excluded from
the present analysis.

\begin{figure}
    \centering
    \includegraphics[width=0.47\linewidth]{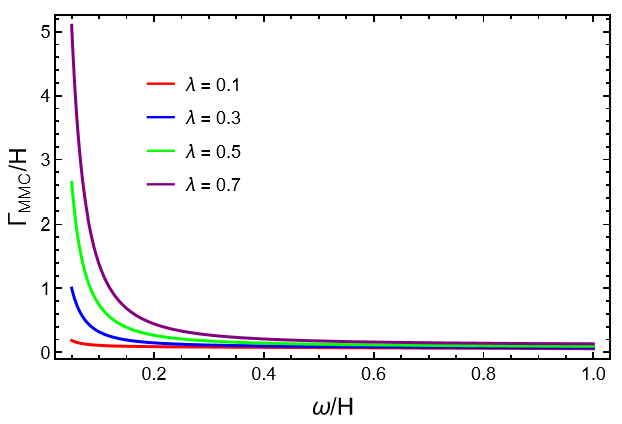}
    \hspace{0.1cm}
    \includegraphics[width=0.50\linewidth]{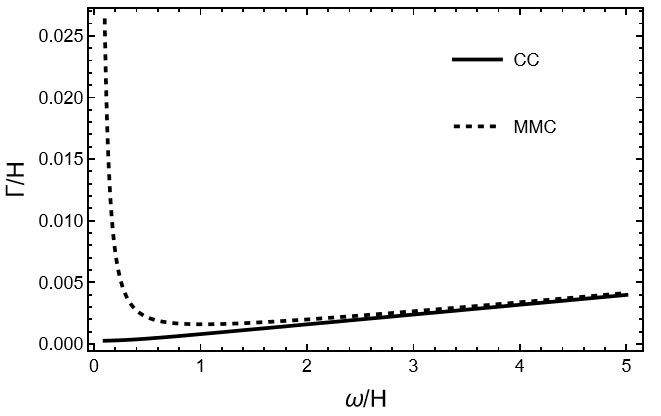}
   \caption{ \it
Dimensionless real asymptotic perturbative decoherence coefficients
for massless minimally coupled (MMC) and conformally coupled (CC)
scalar fields in $(1+3)$-dimensional de Sitter spacetime.
Left: $\Gamma_{\rm MMC}/H$ as a function of the normalized detector
gap $\omega/H$ for representative values of the detector--field
coupling $\lambda$.
The increase at small $\omega/H$ reflects the infrared enhancement
generated by the logarithmic sector of the MMC Wightman function.
Right: comparison of the MMC and CC coefficients at fixed coupling.
For every finite positive detector gap, the MMC coefficient exceeds
the CC coefficient, while the two approach one another as
$\omega/H$ becomes large.
}
    \label{fig:MMC}
\end{figure}

\ref{fig:MMC} illustrates this behavior. The left panel shows
$\Gamma_{\rm MMC}/H$ as a function of $\omega/H$ for representative
values of $\lambda$. The increase at small detector gaps reflects the
infrared enhancement generated by the logarithmic MMC sector. The
right panel compares the CC and MMC coefficients at fixed coupling.
They approach one another for $\omega/H\gg1$, while their difference
becomes increasingly pronounced as the detector gap decreases.

\section{Conclusion and Outlook}
\label{sec:conclusion}

In this work, we investigated the loss of coherence of a two-level
Unruh--DeWitt detector interacting with massless real scalar fields in
$(1+3)$-dimensional de Sitter spacetime. Starting from the
detector--field interaction and working perturbatively to second order
in the coupling, we expressed the evolution of the off-diagonal
element of the reduced detector density matrix in terms of the
positive-frequency Wightman function evaluated along the detector
trajectory. Within the weak-coupling, adiabatic long-interaction, and
secular approximations employed here, the real part of the resulting
coherence kernel defines the asymptotic perturbative decoherence
coefficient considered throughout this work.

For a massless conformally coupled scalar field, the Wightman function
is stationary along a comoving detector trajectory. The corresponding
coefficient is given by~\ref{GammaCC}. In the small-gap limit,
$\omega/H\rightarrow0^+$, the dimensionless coefficient approaches
the finite value
\begin{equation}
\frac{\Gamma_{\rm CC}}{H}
\longrightarrow
\frac{\lambda^2}{4\pi^2},
\end{equation}
whereas for $\omega/H\gg1$ it becomes asymptotically linear in the
dimensionless detector gap $\omega/H$.

The massless minimally coupled field exhibits qualitatively different
behavior owing to the additional logarithmic sector of its Wightman
function. Its correlation function also contains an explicitly
average-time-dependent contribution, denoted by $C(T)$ in our
decomposition. Under the adopted adiabatic prescription, the real part of the
transform of this sector is supported only at zero frequency and
therefore does not contribute to the real asymptotic coefficient at a
fixed finite detector gap $\omega>0$. This conclusion does not imply that the $C(T)$ sector is
absent from the complete finite-duration dynamics, where
switching-dependent boundary contributions may remain. The resulting
MMC coefficient for $\omega>0$ is given by~\ref{GammaMMCfinal}.

Within the weak-coupling, adiabatic long-interaction, and secular
prescription employed here, comparison with the conformally coupled
result shows that
\begin{equation}
\Gamma_{\rm MMC}>\Gamma_{\rm CC},
\qquad
\omega>0,
\end{equation}
as demonstrated in ~\ref{GammaDifference}. The relative
enhancement is summarized by~\ref{GammaRatio},
\begin{equation}
\frac{\Gamma_{\rm MMC}}{\Gamma_{\rm CC}}
=
1+\left(\frac{H}{\omega}\right)^2.
\end{equation}
The distinction between minimal and conformal coupling is therefore
controlled by the ratio $H/\omega$. It is most pronounced when the
detector gap is comparable to or smaller than the de Sitter curvature
scale, while the two coefficients approach one another for
$\omega/H\gg1$.

In the small-gap regime, the conformally coupled coefficient remains
finite, whereas the MMC coefficient exhibits the infrared enhancement
derived in ~\ref{GammaDifferenceSmall}. This enhancement
originates from the logarithmic relative-time sector of the MMC
correlation function. The divergence of the second-order expression
as $\omega/H\rightarrow0^+$ should not be interpreted as an
arbitrarily large physical decoherence rate. Instead, it indicates
that the infrared contribution increasingly dominates the
perturbative coefficient and that fixed-order perturbation theory
eventually loses quantitative reliability near the degenerate limit.
In particular, the accumulated correction must remain perturbatively
small over the observation time. The exactly degenerate case,
$\omega=0$, is therefore excluded from the present analysis.

The principal contribution of this work is the extension of the
conformal-versus-minimal comparison from detector transition
responses to the perturbative evolution of detector coherence. The
half-line coherence kernel makes it possible to identify separately
the roles of the conformal, logarithmic, and explicitly
average-time-dependent sectors of the MMC correlation function.
Within the asymptotic prescription adopted here, the logarithmic
relative-time sector generates the additional real coherence-decay
contribution, whereas the average-time-dependent sector contributes
only to the phase at fixed nonzero detector gap. The resulting
coefficient therefore provides a direct connection between the
infrared structure of the minimally coupled field and the secular
loss of coherence of a localized detector.

A natural extension of the present work is a complete finite-duration
analysis with a specified smooth switching function. Such a treatment
would retain the correlated average- and relative-time integration
domains and determine the full time dependence of the detector
coherence, including switching-dependent boundary contributions from
the $C(T)$ sector. It could also clarify possible non-Markovian and
memory effects that are not captured by the asymptotic coefficient
studied here. Further extensions include massive scalar fields with
general curvature coupling and non-comoving or accelerated detector
trajectories, which would make it possible to investigate how field
mass, curvature coupling, detector motion, and spacetime curvature
jointly influence the loss of quantum coherence.

\begin{acknowledgments}
K.A. and S.K. acknowledge the Department of Physics, School of Advanced
Sciences, Vellore Institute of Technology (VIT), Vellore, for providing
research facilities.
\end{acknowledgments}

\appendix
\section{Evaluation of the Conformally Coupled Contribution}
\label{app:conformally_coupled_residues}

In this appendix, we evaluate the conformally coupled contribution to
the real asymptotic coherence coefficient by relating the two
half-line integrals appearing in the coherence kernel to the full-line
Fourier transform of the Wightman function. The calculation is
performed in the adiabatic long-interaction limit at a fixed positive
detector gap, $\omega>0$.

For a massless conformally coupled scalar field, the Wightman function
along a comoving detector trajectory is
\begin{equation}
iG^+_{\rm CC}(\Delta\tau)
=
-\frac{H^2}{16\pi^2}
\frac{1}{
\left(
\sinh\frac{H\Delta\tau}{2}
-i\epsilon
\right)^2
},
\label{app:GCC}
\end{equation}
where $\epsilon$ implements the usual Wightman boundary-value
prescription. We define the corresponding full-line Fourier transform
by
\begin{equation}
\mathcal F_{\rm CC}(\omega)
=
\int_{-\infty}^{\infty}
d(\Delta\tau)\,
iG_{\rm CC}^{+}(\Delta\tau)
e^{i\omega\Delta\tau}.
\label{app:FCCdef}
\end{equation}

To evaluate ~\ref{app:FCCdef} for $\omega>0$, introduce the
complex variable $z=x+iy$ and define
\begin{equation}
f_{+}(z)
=
\frac{e^{i\omega z}}{
\sinh^{2}\!\left[
\frac{H}{2}(z-i\epsilon)
\right]}.
\label{app:fplus}
\end{equation}
Since
\begin{equation}
e^{i\omega z}
=
e^{i\omega x}e^{-\omega y},
\label{app:exponential_suppression}
\end{equation}
the exponential factor is suppressed in the upper half-plane. The
poles of $f_{+}(z)$ are located at
\begin{equation}
z_n
=
i\epsilon+\frac{2\pi i n}{H},
\qquad
n\in\mathbb Z.
\label{app:polesplus}
\end{equation}
For $\epsilon>0$, an upper-half-plane contour encloses the poles with
$n=0,1,2,\ldots$ in the large-contour limit. In the vicinity of a pole $z=z_n$, we have
\begin{equation}
\sinh\left[
\frac{H}{2}(z-i\epsilon)
\right]
\simeq
(-1)^n\frac{H}{2}(z-z_n),
\label{app:sinh_expansion}
\end{equation}
and therefore
\begin{equation}
\frac{1}{
\sinh^{2}\!\left[
\frac{H}{2}(z-i\epsilon)
\right]}
\simeq
\frac{4}{H^2}
\frac{1}{(z-z_n)^2}.
\label{app:pole_expansion}
\end{equation}
Thus, each singularity is a second-order pole. The corresponding
residue is
\begin{align}
\operatorname{Res}(f_{+},z_n)
&=
\lim_{z\rightarrow z_n}
\frac{d}{dz}
\left[
(z-z_n)^2f_{+}(z)
\right]
\nonumber\\
&=
\frac{4i\omega}{H^2}
e^{i\omega z_n}.
\label{app:resplus}
\end{align}

Before removing the regulator,
\begin{equation}
e^{i\omega z_n}
=
e^{-\omega\epsilon}
e^{-2\pi n\omega/H}.
\label{app:regulated_exponential}
\end{equation}
Taking $\epsilon\rightarrow0^+$ after performing the contour
integration gives
\begin{equation}
e^{i\omega z_n}
\longrightarrow
e^{-2\pi n\omega/H}.
\label{app:exponential_limit}
\end{equation}
The sum of the enclosed residues is consequently
\begin{align}
\sum_{n=0}^{\infty}
\operatorname{Res}(f_{+},z_n)
&=
\frac{4i\omega}{H^2}
\sum_{n=0}^{\infty}
e^{-2\pi n\omega/H}
=
\frac{4i\omega}{H^2}
\frac{1}{1-e^{-2\pi\omega/H}}.
\label{app:ressumplus}
\end{align}

Applying the residue theorem on a suitable sequence of
upper-half-plane contours and then taking the large-contour limit, we
obtain
\begin{align}
\int_{-\infty}^{\infty}
d(\Delta\tau)\,
\frac{e^{i\omega\Delta\tau}}{
\sinh^{2}\!\left[
\frac{H}{2}(\Delta\tau-i\epsilon)
\right]}
&=
2\pi i
\sum_{n=0}^{\infty}
\operatorname{Res}(f_{+},z_n)
=
-\frac{8\pi\omega}{H^2}
\frac{1}{1-e^{-2\pi\omega/H}}.
\label{app:fullplus}
\end{align}
Combining ~\ref{app:fullplus} with the prefactor in~\ref{app:GCC}, we find
\begin{equation}
\mathcal F_{\rm CC}(\omega)
=
\frac{\omega}{2\pi}
\frac{1}{1-e^{-2\pi\omega/H}},
\qquad
\omega>0.
\label{app:FCCplus}
\end{equation}

The corresponding negative-frequency transform is
\begin{equation}
\mathcal F_{\rm CC}(-\omega)
=
\frac{\omega}{2\pi}
\frac{1}{e^{2\pi\omega/H}-1},
\qquad
\omega>0.
\label{app:FCCminus}
\end{equation}
The two Fourier components therefore satisfy
\begin{equation}
\mathcal F_{\rm CC}(-\omega)
=
e^{-2\pi\omega/H}
\mathcal F_{\rm CC}(\omega),
\label{app:detailedbalance}
\end{equation}
which is the expected detailed-balance relation at the de Sitter
temperature $T_{\rm dS}=H/(2\pi)$.

The conformally coupled contribution to the asymptotic coherence
kernel is
\begin{equation}
\mathcal D_{\rm CC}(\omega)
=
\int_{0}^{\infty}d\tau\,
iG_{\rm CC}^{+}(\tau)e^{i\omega\tau}
+
\int_{-\infty}^{0}d\tau\,
iG_{\rm CC}^{+}(\tau)e^{-i\omega\tau}.
\label{app:DCC}
\end{equation}
The Wightman function satisfies
\begin{equation}
iG_{\rm CC}^{+}(-\tau)
=
\left[
iG_{\rm CC}^{+}(\tau)
\right]^*.
\label{app:Wsymmetry}
\end{equation}
Using this relation, the real part of~\ref{app:DCC} can be written as the symmetric combination
\begin{equation}
\operatorname{Re}\mathcal D_{\rm CC}(\omega)
=
\frac{1}{2}
\left[
\mathcal F_{\rm CC}(\omega)
+
\mathcal F_{\rm CC}(-\omega)
\right].
\label{app:relation}
\end{equation}
Thus, the change in the magnitude of the detector coherence is
governed by the symmetric combination of the positive- and
negative-frequency Wightman spectra.

Substituting ~\ref{app:FCCplus} and
\ref{app:FCCminus} into~\ref{app:relation}, we obtain
\begin{align}
\operatorname{Re}\mathcal D_{\rm CC}(\omega)
&=
\frac{\omega}{4\pi}
\left[
\frac{1}{1-e^{-2\pi\omega/H}}
+
\frac{1}{e^{2\pi\omega/H}-1}
\right]
\nonumber\\
&=
\frac{\omega}{4\pi}
\coth\left(
\frac{\pi\omega}{H}
\right),
\qquad
\omega>0.
\label{app:ReDCC}
\end{align}
This reproduces the conformally coupled result quoted in
~\ref{ReDCC_result} of the main text.

\section{Logarithmic Contribution for the Massless Minimally Coupled Field}
\label{app:mmc}

In this appendix, we evaluate the logarithmic relative-time
contribution to the real asymptotic coherence coefficient for the
massless minimally coupled scalar field. The calculation is performed
in the adiabatic long-interaction limit at a fixed positive detector
gap, $\omega>0$. The relative-time-independent sector $C(T)$ is not included here
because, under the adopted prescription, the real part of its
transform is supported only at zero frequency and therefore does not
contribute to the real asymptotic coefficient for $\omega>0$, as
shown in Section~\ref{sec:regulated_CT}.

Along a comoving detector trajectory, the de Sitter-invariant variable
with the Wightman boundary-value prescription is
\begin{equation}
y(\Delta\tau)
=
-4\sinh^2\left[
\frac{H}{2}(\Delta\tau-i\epsilon)
\right],
\qquad
\epsilon\rightarrow0^+.
\label{app:yMMC}
\end{equation}
The logarithmic contribution to the coherence kernel can then be
written as
\begin{equation}
\mathcal D_{\log}(\omega)
=
-\frac{H^2}{8\pi^2}
\int_0^\infty d(\Delta\tau)\,
e^{i\omega\Delta\tau}
\left[
\ln y(\Delta\tau)
+
\ln y(-\Delta\tau)
\right].
\label{B1}
\end{equation}

We introduce the dimensionless variables
\begin{equation}
u
=
\frac{H\Delta\tau}{2},
\qquad
\Omega
=
\frac{2\omega}{H}.
\label{app:variables}
\end{equation}
Using
\begin{equation}
\sinh u
=
\frac{e^u}{2}
\left(1-e^{-2u}\right),
\label{app:sinh_identity}
\end{equation}
the boundary values of the logarithms for $u>0$ are
\begin{align}
\ln y(\Delta\tau)
&=
2u
+
2\ln\left(1-e^{-2u}\right)
+i\pi,
\label{app:log_positive}
\\
\ln y(-\Delta\tau)
&=
2u
+
2\ln\left(1-e^{-2u}\right)
-i\pi.
\label{app:log_negative}
\end{align}
The branch-dependent imaginary terms cancel in their symmetric
combination, giving
\begin{equation}
\ln y(\Delta\tau)
+
\ln y(-\Delta\tau)
=
4u
+
4\ln\left(1-e^{-2u}\right).
\label{B2}
\end{equation}

Substituting~\ref{B2} into~\ref{B1} gives the regulated
expression
\begin{equation}
\mathcal D_{\log}(\omega)
=
-\frac{H}{\pi^2}
\lim_{\eta\rightarrow0^+}
\int_0^\infty du\,
e^{(i\Omega-\eta)u}
\left[
u+\ln\left(1-e^{-2u}\right)
\right],
\label{B3}
\end{equation}
where $\eta>0$ is an adiabatic convergence regulator, distinct from
the Wightman regulator $\epsilon$ appearing in~\ref{app:yMMC}. The limit $\eta\rightarrow0^+$ is taken only
after the integration has been evaluated.

We evaluate the two contributions in~\ref{B3} separately. For
the term linear in $u$, define
\begin{align}
I_u(\Omega)
&=
\lim_{\eta\rightarrow0^+}
\int_0^\infty du\,
u\,e^{(i\Omega-\eta)u}
=
\lim_{\eta\rightarrow0^+}
\frac{1}{(\eta-i\Omega)^2}.
\label{app:Iureg}
\end{align}
As a distribution, this expression is
\begin{equation}
I_u(\Omega)
=
-\mathcal P\left(\frac{1}{\Omega^2}\right)
-i\pi\delta'(\Omega),
\label{Iu}
\end{equation}
where $\mathcal P$ denotes the principal-value distribution. Since the
present calculation is restricted to $\Omega>0$, the distributional
term supported at $\Omega=0$ does not contribute. Therefore,
\begin{equation}
I_u(\Omega)
=
-\frac{1}{\Omega^2},
\qquad
\Omega>0.
\label{B4}
\end{equation}

For the remaining contribution, we use the convergent series
\begin{equation}
\ln\left(1-e^{-2u}\right)
=
-\sum_{n=1}^{\infty}
\frac{e^{-2nu}}{n},
\qquad
u>0.
\label{app:logseries}
\end{equation}
Because this term is integrable both at $u=0$ and as
$u\rightarrow\infty$, its regulator may be removed after integration.
We consequently obtain
\begin{align}
I_{\log}(\Omega)
&=
\lim_{\eta\rightarrow0^+}
\int_0^\infty du\,
e^{(i\Omega-\eta)u}
\ln\left(1-e^{-2u}\right)
\nonumber\\
&=
-\sum_{n=1}^{\infty}
\frac{1}{n}
\int_0^\infty du\,
e^{-(2n-i\Omega)u}
\nonumber\\
&=
-\sum_{n=1}^{\infty}
\frac{1}{n(2n-i\Omega)}.
\label{B5}
\end{align}

Using the partial-fraction identity
\begin{equation}
\frac{1}{n(2n-i\Omega)}
=
-\frac{1}{i\Omega}
\left[
\frac{1}{n}
-
\frac{1}{n-i\Omega/2}
\right],
\label{app:partialfraction}
\end{equation}
together with the defining series of the digamma function,
\begin{equation}
\sum_{n=1}^{\infty}
\left[
\frac{1}{n}
-
\frac{1}{n+z}
\right]
=
\psi(1+z)+\gamma,
\label{app:digammaseries}
\end{equation}
where $\gamma$ is the Euler--Mascheroni constant, we find
\begin{equation}
I_{\log}(\Omega)
=
-\frac{i}{\Omega}
\left[
\psi\left(
1-\frac{i\Omega}{2}
\right)
+\gamma
\right].
\label{B6}
\end{equation}

Combining~\ref{B4} and \ref{B6} in~\ref{B3}, we obtain
\begin{equation}
\mathcal D_{\log}(\omega)
=
\frac{H}{\pi^2\Omega^2}
+
\frac{iH}{\pi^2\Omega}
\left[
\psi\left(
1-\frac{i\Omega}{2}
\right)
+\gamma
\right].
\label{B7}
\end{equation}
Restoring $\Omega=2\omega/H$ gives
\begin{equation}
\mathcal D_{\log}(\omega)
=
\frac{H^3}{4\pi^2\omega^2}
+
\frac{iH^2}{2\pi^2\omega}
\left[
\psi\left(
1-\frac{i\omega}{H}
\right)
+\gamma
\right],
\qquad
\omega>0.
\label{Bfinal}
\end{equation}

Only the real part of the kernel contributes to the asymptotic change
in the magnitude of the detector coherence. Since $\gamma$ is real,~\ref{Bfinal} yields
\begin{equation}
\operatorname{Re}\mathcal D_{\log}(\omega)
=
\frac{H^3}{4\pi^2\omega^2}
-
\frac{H^2}{2\pi^2\omega}
\operatorname{Im}
\psi\left(
1-\frac{i\omega}{H}
\right).
\label{app:ReDlog}
\end{equation}
For $x>0$, the imaginary part of the digamma function satisfies
\begin{equation}
\operatorname{Im}\psi(1-ix)
=
-\frac{\pi}{2}\coth(\pi x)
+\frac{1}{2x}.
\label{app:digammaidentity}
\end{equation}
Setting $x=\omega/H$ in~\ref{app:digammaidentity} gives
\begin{align}
\operatorname{Re}\mathcal D_{\log}(\omega)
=
\frac{H^3}{4\pi^2\omega^2}
+
\frac{H^2}{4\pi\omega}
\coth\left(
\frac{\pi\omega}{H}
\right)
-
\frac{H^3}{4\pi^2\omega^2}
=
\frac{H^2}{4\pi\omega}
\coth\left(
\frac{\pi\omega}{H}
\right).
\label{app:ReDlog_reduction}
\end{align}
Thus, the terms proportional to $H^3/\omega^2$ cancel exactly, leaving
\begin{equation}
\operatorname{Re}\mathcal D_{\log}(\omega)
=
\frac{H^2}{4\pi\omega}
\coth\left(
\frac{\pi\omega}{H}
\right),
\qquad
\omega>0.
\label{app:ReDlogfinal}
\end{equation}
This is the logarithmic relative-time contribution used in~\ref{ReDlogsimplified} of the main text.

%\bibliographystyle{cas-model2-names}
%\bibdata{outputNotes,outputNotes : ,cas-refs}

\end{document}